# Sinusoidal Frequency Doublers Circuit With Low Voltage ± 1.5 Volt CMOS Inverter

Bancha Burapattanasiri

Department of Electronic and Telecommunication Engineering, Engineering Collaborative Research Center

Faculty of Engineering, Kasem Bundit University

Bangkok, Thailand 10250
.

*Abstract*—This paper is present sinusoidal frequency doublers circuit with low voltage ± 1.5 volt CMOS inverter. Main structure of circuit has three parts that is CMOS inverter circuit, differential amplifier circuit, and square root circuit. This circuit has designed to receive input voltage and give output voltage use few MOS transistor, easy to understand, non complex of circuit, high precision, low error and low power. The Simulation of circuit has MOS transistor functional in active and saturation period. PSpice programmed has used to confirmation of testing and simulation.

*Keywords-component; sinusoidal frequency doublers, low voltage,CMOS inverter.*

## I. INTRODUCTION

Sinusoidal frequency doublers is popular in telecommunication for example using instrument processing, or circuit analysis in analog processing [1-2]. The normally, sinusoidal frequency doublers has be the characteristic of tune LC circuit or analog multiplier circuit. A lot of researches presented has the narrow frequency operation period and non suitable for establish of integrated circuit, so the circuit development by multiplier circuit and sinusoidal frequency doublers used op-amp to function, then it able to charge the limited of tune LC circuit, and able to establish integrated circuit too. However, the circuits still have op-amp then the circuits still have big size, high loss of power supply, and used a lot of device. So in this paper is present new choice of sinusoidal frequency doublers and suitable for establish high integrated circuit. Because of the circuit has designed easy to understand, noncomplex, and MOS transistor functional in active and saturation regions for compound to CMOS inverter circuit, Differential amplifier circuit and square-rooter circuit. The new circuits is of MOS transistor, but still have high efficiency, low power supply ± 1.5 Volt, high precision low error and low power. From the purpose of research doesn't want error with circuit, so setting all K of MOS transistors is equal.

## II. DESIGNATION AND FUNCTIONAL

The sinusoidal frequency doublers circuit with low voltage ± 1.5 volt has three parts of circuit. In the first part is CMOS inverter circuit has functional to invert input signal high speed, low errors and high precision. The second part is differential amplifier circuit has function to squares rule of circuit. The third part is square-rooter circuit has functional to squared of integrated differential amplifier circuit by the relation of MOS transistor in active and saturation period, when the functional three part of circuit put together is able to write diagram box as figure 1. For more understand in this researcher we are separate the part of circuit operation as following.

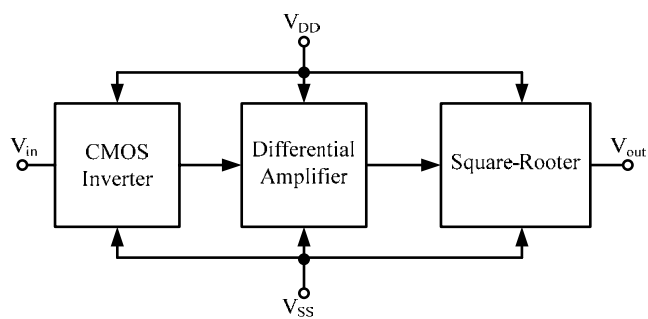

Figure 1. Diagram is show sinusoidal frequency doublers circuit with low voltage ± 1.5 volt CMOS inverter.

### A. CMOS Iverter Crcuit

From the figure 2 show CMOS inverter circuit component by $M_1$, $M_2$, $M_3$ and $M_4$. So, Sending input signal has to positive and negative phase entrance pin-gate $M_1$ and $M_2$ it will two MOS transistor serration working. To be a result of CMOS inverter designed working in MOS transistor saturation. So we can computable output–input relation of CMOS inverter by

$$I_{DM1} = I_{DM3} \text{ And } I_{DM2} = I_{DM4} \quad (1)$$

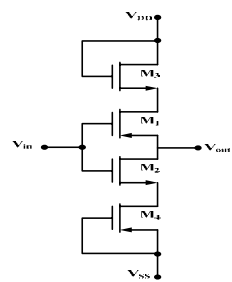

Figure 2. CMOS inverter circuit

An equation current of MOS transistor saturation working is

$$I_D = K(V_{GS} - V_T)^2 ; (V_{GS} - V_T) \leq V_{DS} \quad (2)$$

When $\quad K = \dfrac{\varepsilon_{ins}\varepsilon_O \mu_n}{D}$ and $\beta = K(W/L)$ (3)

Whence (1) equation and CMOS inverter circuit value to result is

$$I_{DM1} = -I_{DM2} \quad (4)$$

And $\quad I_{DM1} = \dfrac{-\beta_p}{2}(V_{in} - V_{DD} - V_{tp})^2 \quad (5)$

$$I_{DM2} = \dfrac{\beta_n}{2}(V_{in} - V_{tn})^2 \quad (6)$$

Then $\quad V_{in} = -\left(\dfrac{V_{DD} + V_{tp} + V_{tn}(\beta_n/\beta_p)^{1/2}}{1 + (\beta_n/\beta_p)^{1/2}}\right) \quad (7)$

If a symmetry point of circuit in equation (8)

$$V_{in} = V_{out} = V_{DD}/2 \quad (8)$$

By $\quad \beta_n = \beta_p \text{ and } V_{tn} = -V_{tp} \quad (9)$

So $\quad V_{out} = -V_{in} \quad (10)$

*B. Differential Amplifier Circuit*

From the figure 3 is show a differential amplifier circuit component $M_5$ and $M_6$. Functional of circuit is working when $M_5$ and $M_6$ receive signal from CMOS inverter circuit, and to setting MOS transistor working by square formulation. The relation $M_5$ and $M_6$ current able to show [5]

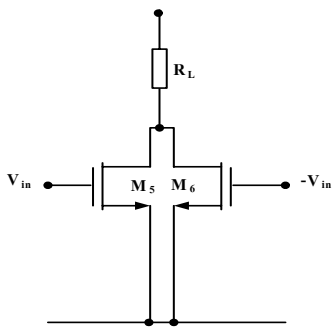

Figure 3. Differential amplifier circuit

$$I_{RL} = I_{DM5} + I_{DM6} \quad (11)$$

By $\quad I_{DM5} = K_5(V_{in} - V_{ss} - V_T)^2 \quad (12)$

And $\quad I_{DM6} = K_6(-V_{in} - V_{SS} - V_T)^2 \quad (13)$

Instead of an equation (12) and (13) in to (11) then the result is

$$I_{RL} = \left[K_5(V_{in} - V_{SS} - V_T)^2\right] + \left[K_6(-V_{in} - V_{SS} - V_T)^2\right] \quad (14)$$

From the designing is setting $K_5 = K_6 = K$

So $\quad I_{RL} = 2K\left[(V_{in})^2 + (V_{SS} + V_T)^2\right] \quad (15)$

*C. Square-Rooter Circuit*

From the figure 4 is show a square-rooter circuit component $M_7$ and $M_8$. The circuit designing is setting $M_7$ work in saturation period and $M_8$ work in non-saturation period. The circuit relation show as an equation current [6]

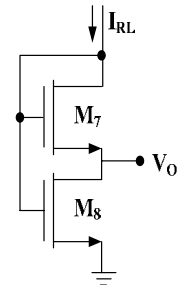

Figure 4. Square-rooter circuit

$$I_{DM7} = K(V_{DM7} - V_O - V_T)^2 \quad (16)$$

$$I_{DM8} = K\left[(V_{DM8} - V_T)V_O - \dfrac{V_O^2}{2}\right] \quad (17)$$

From figure 4 if $I_{DM7} = I_{DM8}$ so, the relation Voltage output Circuit computable by

$$V_{DM8} = \sqrt{\dfrac{I_{DM7}}{K}} + V_O + V_T \quad (18)$$

When instead result of an equation (18) to (17) the new result is

$$I_{DM7} = I_{DM8} = K\left[\left(\sqrt{\dfrac{I_{DM7}}{K}}V_O + V_T - V_T\right)V_O - \dfrac{V_O^2}{2}\right] \quad (19)$$

$$K\frac{V_O^2}{2}+K\sqrt{\frac{I_{DM7}}{K}}V_0^2-I_{DM7}=0 \quad (20)$$

From equation (20) is able mathematic calculation to finding $V_O$ then the result is square root of drain current as (21)

$$V_O=\frac{0.732}{\sqrt{K}}\sqrt{I_{DM7}} \quad (21)$$

*D. The Completely Sinusoidal Frequency Doublers Circuit With Low Voltage ± 1.5 volt CMOS Inverter.*

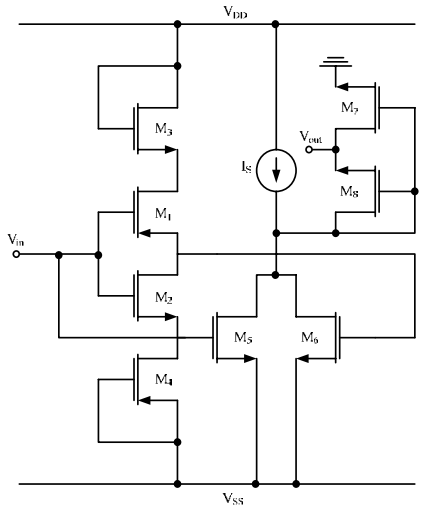

Figure 5. The completely sinusoidal frequency doublers circuit with low voltage ± 1.5 volt CMOS inverter.

From the function of three circuits, when we put gathers, so the new circuit is the completely sinusoidal frequency doublers circuit with low voltage ± 1.5 volt, it has high precision, low error, used few MOS transistor, low power. When bring equation (10) of CMOS inverter circuit to analyze by setting $V_{in} = V_m \sin \omega t$ the output CMOS inverter result is $-V_{in}$. While, the differential amplifier circuit integrated to inverter CMOS circuit, and square-rooter circuit integrated to differential amplifier circuit. So, the new result has related to all result of circuit, it is able to write the output circuit equation as

$$V_O=0.732\sqrt{\frac{1}{K}(2KV_{in}^2)} \quad (22)$$

$$V_O=1.035\sqrt{V_{in}^2} \quad (23)$$

When bring $V_{in} = V_m \sin \omega t$ instead in equation (23) and analyze by trigonometric function relate is

$$Sin^2\theta=1-Cos^2\theta \quad (24)$$

If the power series of the from

$$\sqrt{1+X}=1+\left(\frac{1}{2}\right)x+\left(\frac{1}{8}\right)x^2+.....are\ employed \quad (25)$$

Then the equation (23) can be rewritten as

$$V_O=1.035(V_{DC}+V_mCos^2\omega t) \quad (26)$$

III. SIMULATION AND MEASUREMENT RESULT

From the completely sinusoidal frequency doublers circuit with low voltage ± 1.5 volt CMOS inverter is abler to confirmation the efficient of simulation circuit by PSpice programmed to analyze and the testing by send input voltage signal as equation (27) setting $V_{DD}$ = 1.5 Volt, $V_{SS}$ = -1.5 Volt and W/L = 1.5/0.15 μm, by sending input voltage signal at equation (27), and to setting W/L = 1.5/0.15 μm, then the result is output signal as figure 6

$$V_{in}=0.1\ Sin\ 2000\pi t \quad (27)$$

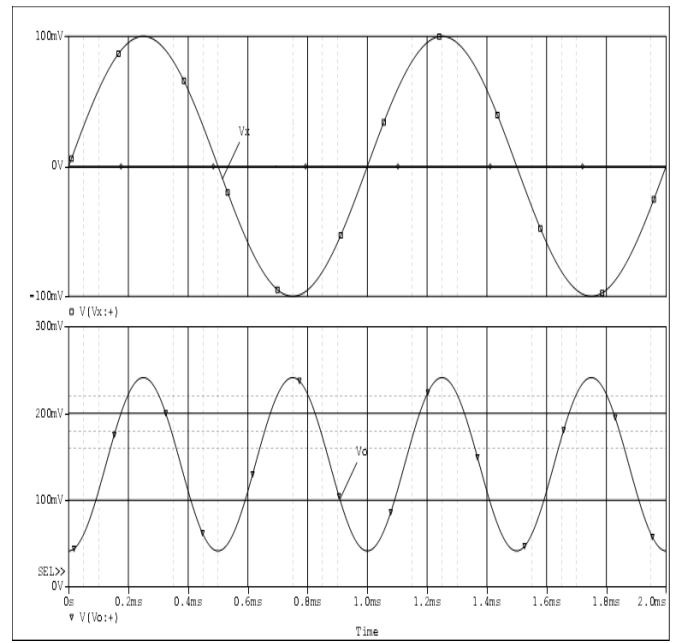

Figure 6. The result from sending input signal as (27)

IV. CONCLUSION

From the sinusoidal frequency doublers circuit with low voltage ± 1.5 volt CMOS inverter has presented that show noncomplex of working function, dissipation of current source, and few MOS transistor, operating at input and output in voltage mode, high precision, low error. The efficient simulation circuit is able confirm by PSpice program as presentation principle.

## APPENDIX

The parameters used in simulation are 0.5 μm CMOS Model obtained through MIETEC [10] as listed in Table I. For aspect ratio (W/L) of MOS transistors used are as follows: 1.5 μm / 0.15 μm for all NMOS transistors; 1.5 μm / 0.15 μm for all PMOS transistors.

TABLE I.  CMOS MODEL USED IN THE SIMULATION

```
-------------------------------------------------------------------------------
.MODEL CMOSN NMOS LEVEL = 3 TOX = 1.4E-8 NSUB = 1E17
GAMMA = 0.5483559 PHI = 0.7 VTO = 0.7640855 DELTA = 3.0541177
UO = 662.6984452 ETA = 3.162045E-6 THETA = 0.1013999
KP = 1.259355E-4 VMAX = 1.442228E5 KAPPA = 0.3 RSH = 7.513418E-3
NFS = 1E12 TPG = 1 XJ = 3E-7 LD = 1E-13 WD = 2.334779E-7
CGDO = 2.15E-10 CGSO = 2.15E-10 CGBO = 1E-10 CJ = 4.258447E-4
PB = 0.9140376 MJ = 0.435903 CJSW = 3.147465E-10 MJSW = 0.1977689

.MODEL CMOSP PMOS LEVEL = 3 TOX = 1.4E-8 NSUB = 1E17
GAMMA = 0.6243261 PHI = 0.7 VTO = -0.9444911 DELTA = 0.1118368
UO = 250 ETA = 0 THETA = 0.1633973 KP = 3.924644E-5 VMAX = 1E6
KAPPA = 30.1015109 RSH = 33.9672594 NFS = 1E12 TPG = -1 XJ = 2E-7
LD = 5E-13 WD = 4.11531E-7 CGDO = 2.34E-10 CGSO = 2.34E-10
CGBO = 1E-10 CJ = 7.285722E-4 PB = 0.96443 MJ = 0.5
CJSW = 2.955161E-10 MJSW = 0.3184873
-------------------------------------------------------------------------------
```

## ACKNOWLEDGMENT

The researchers, we are thank you very much to our parents, who has supporting everything to us. Thankfully to all of professor for knowledge and a consultant, thank you to Miss Suphansa Kansa-Ard for her time and supporting to this research. The last one we couldn't forget that is Kasem Bundit University, Engineering Faculty for supporting and give opportunity to our to development in knowledge and research, so we are special thanks for everything.

## AUTHORS PROFILE

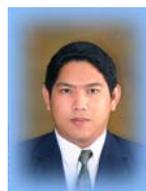

Mr.Bancha Burapattanasiri received the bleacher degree in electronic engineering from Kasem Bundit University in 2002 and master degree in Telecommunication Engineering, from King Mongkut's Institute of Technology Ladkrabang in 2008. He is a lecture of Electronic and Telecommunication Engineering, Faculty of Engineering, Kasem Bundit University, Bangkok, Thailand. His research interests analog circuit design, low voltage, high frequency and high-speed CMOS technology.